%% file: SUS-16-037_temp.tex
\pdfoutput=1

\documentclass[11pt,twoside,a4paper,cmspaper,final,collab]{cms-tdr}
\def\svnVersion{430331P}\def\cmsCernNoTag{CERN-EP-2017-088}\def\cmsCernDate{\today}\def\cmsMessage{Published in Physical Review Letters as \href{http://dx.doi.org/10.1103/PhysRevLett.119.151802}{\doi{10.1103/PhysRevLett.119.151802.}}}

\begin{document}\cmsNoteHeader{SUS-16-037}

\hyphenation{had-ron-i-za-tion}
\hyphenation{cal-or-i-me-ter}
\hyphenation{de-vices}
\RCS$Revision: 420218 $
\RCS$HeadURL: svn+ssh://svn.cern.ch/reps/tdr2/papers/SUS-16-037/trunk/SUS-16-037.tex $
\RCS$Id: SUS-16-037.tex 420218 2017-08-08 22:20:55Z richman $
\newlength\cmsFigWidth
\ifthenelse{\boolean{cms@external}}{\setlength\cmsFigWidth{0.98\columnwidth}}{\setlength\cmsFigWidth{0.9\textwidth}}
\newlength\cmsFigWidthii
\ifthenelse{\boolean{cms@external}}{\setlength\cmsFigWidthii{0.49\columnwidth}}{\setlength\cmsFigWidthii{0.45\textwidth}}
\ifthenelse{\boolean{cms@external}}{\providecommand{\cmsLeft}{top\xspace}}{\providecommand{\cmsLeft}{left\xspace}}
\ifthenelse{\boolean{cms@external}}{\providecommand{\cmsRight}{bottom\xspace}}{\providecommand{\cmsRight}{right\xspace}}
\ifthenelse{\boolean{cms@external}}{\providecommand{\CL}{\ensuremath{\text{C.L.}}\xspace}}{\providecommand{\CL}{\ensuremath{\mathrm{CL}}\xspace}}

\newcommand{\largr}{large-\ensuremath{R}\xspace}
\newcommand{\smalr}{small-\ensuremath{R}\xspace}
\newcommand{\MJ}{\ensuremath{M_J}\xspace}
\newcommand{\njets}{\ensuremath{N_{\text{jets}}}\xspace}
\newcommand{\nb}{\ensuremath{N_{\PQb}}\xspace}
\newcommand{\ST}{\ensuremath{S_{\mathrm{T}}}\xspace}
\newcommand{\mGlu}{\ensuremath{m(\PSg)}\xspace}
\newcommand{\mLSP}{\ensuremath{m(\PSGczDo)}\xspace}
\newcommand{\mTt}{\ensuremath{M_{\mathrm{T}2}}\xspace}
\newcommand{\mt}{\ensuremath{m_{\mathrm{T}}}\xspace}
\newcommand\zjets{\ensuremath{\Z\text{+jets}}\xspace}
\newcommand\wjets{\ensuremath{\PW\text{+jets}}\xspace}
\newcommand\ttjets{\ensuremath{\ttbar\text{+jets}}\xspace}
\newcommand{\ptvecell}{\ensuremath{\vec{p}^{\kern1pt\ell}}\xspace}
\newcommand{\ptvecTv}{\ensuremath{\vec{p}_{\mathrm{T}}^{\kern2pt v}}\xspace}

\cmsNoteHeader{SUS-16-037}
\title{Search for supersymmetry in \texorpdfstring{$\Pp\Pp$}{pp} collisions at \texorpdfstring{$\sqrt{s} = 13\TeV$}{sqrt(s) = 13 TeV} in the single-lepton final state using the sum of
masses of large-radius jets}

\date{\today}

\abstract{Results are reported from a search for supersymmetric particles in proton-proton collisions in the final state with a single lepton; multiple jets, including at least one \PQb-tagged jet; and large missing transverse momentum. The search uses a sample of proton-proton collision data at $\sqrt{s}=13\TeV$ recorded by the CMS experiment at the LHC, corresponding to an integrated luminosity of 35.9\fbinv. The observed event yields in the signal regions are consistent with those expected from standard model backgrounds. The results are interpreted in the context of simplified
models of supersymmetry involving gluino pair production, with gluino decay into either on- or off-mass-shell top squarks. Assuming that the top squarks decay into a top quark plus a stable, weakly interacting neutralino, scenarios with gluino masses up to about 1.9\TeV are excluded at 95\% confidence level for neutralino masses up to about 1\TeV.}

\hypersetup{%
pdfauthor={CMS Collaboration},%
pdftitle={Search for supersymmetry in pp collisions at sqrt(s) = 13 TeV in the single-lepton final state using the
sum of masses of large-radius jets},%
pdfsubject={CMS},%
pdfkeywords={CMS, physics, supersymmetry}}

\maketitle

A central goal of the physics program of the CMS experiment at the CERN LHC~\cite{1748-0221-3-08-S08001}
is the search for new particles and phenomena beyond the standard model (SM), in particular, for supersymmetry
(SUSY)~\cite{Ramond:1971gb,Golfand:1971iw,Neveu:1971rx,Volkov:1972jx,Wess:1973kz,Wess:1974tw,Fayet:1974pd,Nilles:1983ge}.
During 2016, CMS recorded a data sample of proton-proton collisions at a center-of-mass
energy of 13\TeV, corresponding to an integrated luminosity of 35.9\fbinv,
significantly extending the sensitivity to the production of new heavy particles.
The search described here focuses on a generically important experimental signature that is
also strongly motivated by SUSY phenomenology. This signature includes
a single lepton (an electron or a muon);
several jets, arising from the hadronization of energetic quarks and gluons;
at least one \PQb-tagged jet, indicative of processes involving third generation quarks;
and, finally, $\ptvecmiss$, the missing momentum in the direction transverse to the beam. A large value of $\ptmiss \equiv \abs{\ptvecmiss}$
can arise from the production of high momentum, weakly interacting
particles that escape detection. Searches for SUSY in the single-lepton final state have been performed by both
ATLAS and CMS at $\sqrt{s}=7$ and 8\TeV~\cite{ATLAS:2011ad,Aad:2015mia,Chatrchyan:2011qs,Chatrchyan:2013iqa} and
at $\sqrt{s} = 13$\TeV~\cite{Aad:2016eki,Aad:2016qqk,Khachatryan:2016uwr,Khachatryan:2016xdt}.
The present analysis, which introduces extended binning and other improvements,
is based largely on methodologies described in detail in Ref.~\cite{Khachatryan:2016uwr}, which include
the use of large-radius jets and related kinematic variables.

In models based on SUSY, new particles are introduced such that all fermionic
(bosonic) degrees of freedom in the SM are paired with corresponding bosonic (fermionic)
degrees of freedom in the extended theory. The discovery of a Higgs boson with low
mass~\cite{Aad:2012tfa,Chatrchyan:2012ufa,Chatrchyan:2013lba,Khachatryan:2014jba,Aad:2014aba,Aad:2015zhl}
provides a key motivation for SUSY. Stabilizing the Higgs boson mass at a low value, without invoking extreme
fine tuning of parameters, is a major theoretical challenge, referred to
as the gauge hierarchy problem~\cite{tHooft:1979bh,Witten:1981nf,Dine:1981za,Dimopoulos:1981au,Dimopoulos:1981zb,Kaul:1981hi}.
This stabilization can be achieved in so-called natural SUSY models~\cite{1988NuPhB.306...63B,Dimopoulos:1995mi,Barbieri:2009ev,Papucci:2011wy,Feng:2013pwa},
in which several of the SUSY partners are constrained to be light~\cite{Papucci:2011wy}: the
top squarks, $\PSQt_{\mathrm{L}}$
and $\PSQt_{\mathrm{R}}$, which have the same electroweak couplings as the left- (L) and right- (R) handed top
quarks, respectively; the bottom squark with L-handed couplings ($\PSQb_{\mathrm{L}}$); the gluino ($\PSg$); and the
higgsinos ($\sHig$).
This search targets gluino pair production, which has a relatively large cross section for a given mass,
with gluino decay $\PSg\to\ttbar\PSGczDo$. This process can arise from
$\PSg\to\PSQt_1\cPaqt$, where the lighter top squark mass eigenstate $\PSQt_1$ is produced either on or off mass shell. The symbol
$\PSGczDo$ denotes the lightest neutralino, an electrically neutral mass eigenstate that is in general
a mixture of the higgsinos and electroweak gauginos.
In $R$-parity conserving SUSY models~\cite{Farrar:1978xj,Martin:1997ns}
in which the $\PSGczDo$ is the lightest supersymmetric particle (LSP), the $\PSGczDo$ is stable and can, in principle, account
for some or all of the astrophysical dark matter~\cite{Zwicky:1933gu,Rubin:1970zza,Olive:2016xmw}.
The scenario with off-mass-shell top squarks
is denoted as T1tttt~\cite{Chatrchyan:2013sza} in simplified model scenarios~\cite{bib-sms-2,bib-sms-3,bib-sms-4}.
In natural SUSY models, the top squark is typically lighter than the gluino,
so we also search for scenarios with on-shell top squarks, denoted as T5tttt.

Simulated event samples for SM background processes are used to determine correction factors, typically near
unity, that are used in conjunction with observed event yields in control regions to determine the SM
background contribution in the signal regions.  The production of \ttjets, \wjets, \zjets, and quantum
chromodynamics (QCD) multijet events is simulated with the Monte Carlo (MC)
generator \MGvATNLO~2.2.2~\cite{Alwall:2014hca}, with parton distribution functions taken from NNPDF
3.0~\cite{Ball:2014uwa}.  Details on the simulated SM background samples, including other processes with
smaller contributions (single top quark, $\ttbar+$bosons, diboson, and $\ttbar\ttbar$ production) are given in
Ref.~\cite{Khachatryan:2016uwr}. The detector simulation is performed
with \GEANTfour~\cite{Agostinelli:2002hh}.  Simulated event samples for SUSY signal models, used to determine
the selection efficiency for signal events, are generated with \MGvATNLO~2.2.2 with up to two additional
partons at leading order accuracy and are normalized to cross sections based on
Ref.~\cite{Borschensky:2014cia}.  Because of the large number of mass hypotheses examined in this analysis, the
detector simulation in this case is performed with the CMS fast simulation package~\cite{Abdullin:2011zz}.

Two T1tttt benchmark models are used to illustrate typical signal behavior.  The
T1tttt(1800,100) model, which we refer to as a noncompressed-spectrum model (NC), has
$\mGlu=1800$\GeV, $\mLSP=100$\GeV, and a cross section of 2.8\unit{fb}, and
corresponds to a scenario with a large gluino-neutralino mass splitting.
The T1tttt(1400,1000) model, with $\mGlu=1400$\GeV,
$\mLSP=1000$\GeV, and a cross section of 25\unit{fb},
corresponds to a scenario with a small gluino-neutralino mass splitting and is referred to as
a compressed-spectrum model (C).

The data were recorded with the CMS detector~\cite{Chatrchyan:2008zzk}, which is
constructed around a superconducting solenoid of 6\unit{m} diameter,
providing a magnetic field of 3.8\unit{T}. Within the solenoid volume are the charged particle tracking
systems, composed of silicon-pixel and silicon-strip detectors, and the calorimeter systems,
consisting of a lead tungstate crystal electromagnetic calorimeter (ECAL) and a brass and
scintillator hadron calorimeter. Muons are identified and measured by gas-ionization
detectors embedded in the magnetic flux-return yoke outside the solenoid.
Events were selected using several triggers~\cite{Khachatryan:2016bia} that require either large $\ptmiss$
or a single lepton (an electron or a muon), with and without significant hadronic activity.
The trigger efficiency is measured in data for our analysis requirements to be nearly 100\%.

Event reconstruction proceeds from particles identified by the particle-flow (PF)
algorithm~\cite{Sirunyan:2017ulk}, which uses information
from the tracker, calorimeters, and muon systems to identify PF candidates as
electrons, muons, charged or neutral hadrons, or photons.
Electrons are reconstructed by associating a charged-particle track with ECAL
superclusters~\cite{Khachatryan:2015hwa}. The resulting candidate electrons are
required to have transverse momentum $\pt>20\GeV$ and pseudorapidity $\abs{\eta}<2.5$, and to satisfy identification
criteria designed to reject light-parton jets and photon conversions.
Muons are reconstructed by associating tracks in the muon system with
those found in the silicon tracker~\cite{Chatrchyan:2012xi}. Muon candidates
are required to satisfy $\pt>20\GeV$ and $\abs{\eta}<2.4$.
To select leptons from $\PW$ boson decays,
leptons are required to be isolated from other PF candidates.
Isolation is quantified using an optimized version~\cite{Khachatryan:2016uwr} of the mini-isolation variable
originally suggested in Ref.~\cite{Rehermann:2010vq}, in which the transverse energy of
the particles within a cone around the lepton momentum vector is computed
using a cone size that decreases as $1/\pt^{\ell}$, where $\pt^{\ell}$ is the transverse momentum of the lepton.

To suppress dilepton backgrounds, we veto events that contain a broader category
of candidates for the second lepton, referred to as veto tracks.
These include two categories of charged-particle tracks:
isolated leptons satisfying looser identification criteria than lepton candidates, including a
relaxed momentum requirement, $\pt>10$\GeV, and isolated charged-hadron PF candidates,
which must satisfy $\pt>15$\GeV and $\abs{\eta}<2.5$.
In either case, the charge of the veto track must be opposite to that of the lepton candidate in the event.
To maintain a high selection efficiency for signal events, lepton veto tracks must satisfy
a requirement on the quantity~\cite{Lester:mt2,Barr:2003rg}
$\mTt(\ptvecell,\ptvecTv,\ptvecmiss) < 80$\GeV and hadronic veto tracks must satisfy
$\mTt(\ptvecell,\ptvecTv,\ptvecmiss) < 60$\GeV, where $v$ refers to the veto track.

Charged and neutral PF candidates are
clustered into jets using the anti-$\kt$ algorithm~\cite{Cacciari:2008gp}
with radius parameter $R=0.4$, as implemented in the \FASTJET package~\cite{Cacciari:2011ma}.
Jets are required to satisfy $\pt>30$\GeV and $\abs{\eta}\le 2.4$.
Additional details and references are given in Ref.~\cite{Khachatryan:2016uwr}
on the $\pt$- and $\eta$-dependent jet energy calibration~\cite{Chatrchyan:2011ds}, the jet
identification requirements, and the subtraction of the energy contribution to the
jet $\pt$ from multiple proton-proton interactions
from the same or neighboring beam crossings (pileup)~\cite{Cacciari:2007fd}.
A subset of the jets are tagged as originating from \PQb quarks using the
combined secondary vertex algorithm~\cite{Chatrchyan:2012jua,CMS:2016kkf}.

We further cluster the jets with $R=0.4$ (\smalr jets), including those associated with
isolated leptons, into $R=1.4$ (\largr) jets using the anti-$\kt$ algorithm.
The masses, $m(J_i)$, of the \largr jets reflect the \pt spectrum
and multiplicity of the clustered objects, as well as their
angular spread.
The variable \MJ is defined as the sum of all
\largr jet masses:
  $\MJ \!\!= \!\!\sum_{J_i = {\text{large-}R\text{ jets}}} m(J_i)$.
For \ttbar events with a small contribution from initial-state radiation (ISR), the \MJ distribution has an
approximate cutoff at $2m_{\PQt}$.
In contrast, the \MJ distribution for signal events extends to larger values because of the presence
of multiple top quarks in the decay chain.
The presence of a significant amount of ISR generates a high-$\MJ$ tail in the \ttbar
background, producing the main source of background in the analysis.

The missing transverse momentum, $\ptvecmiss$, is defined as
the negative vector sum of the transverse momenta of all PF
candidates.
To separate backgrounds characterized by the presence of a single
$\PW$ boson decaying leptonically, but without any other source of $\ptmiss$,
we use the transverse mass
  $\mt = \sqrt{\smash[b]{2\pt^{\ell}\ptmiss[1-\cos(\Delta\phi_{\ell, \ptvecmiss } )]}}$,
where $\Delta\phi_{\ell,\ptvecmiss}$ is the difference between the
azimuthal angles of $\pt^{\ell}$ and $\ptvecmiss$.
The quantity \HT is defined as the scalar sum of the transverse momenta of all
the \smalr jets passing the selection, while $\ST=\HT + \pt^{\ell}$.

We select events with exactly one isolated charged lepton (an electron or a muon), no veto tracks,
$\ST>500$\GeV, $\ptmiss>200$\GeV, and at least six small-$R$ jets, at least one of which is \PQb tagged.
After this set of requirements, referred to as the {\it baseline selection}, about
80\% of the SM background arises from \ttbar production. The contributions from events with a single
top quark or a \PW~boson in association with jets are each about 6--8\%; much of the remainder
arises from events with a $\ttbar$ pair produced in association with a vector boson.  After applying the
baseline selection, the background from QCD multijet events is negligible.

The analysis is performed using four regions in the $\MJ$-$\mt$ plane: three control regions (CR) and one signal region (SR):
\begin{itemize}
\item R1 (CR): $\mt\le 140$ \GeV,\, $250 \le \MJ\le 400$ \GeV,
\item R2 (CR): $\mt\le 140$ \GeV,\, $\MJ> 400$ \GeV,
\item R3 (CR): $\mt>140$ \GeV,\, $250 \le \MJ\le 400$ \GeV,
\item R4 (SR): $\mt>140$ \GeV,\, $\MJ> 400$ \GeV.
\end{itemize}
All four regions are divided in bins of \ptmiss, forming three largely independent $\MJ$-$\mt$ planes:
\begin{itemize}
\item three \ptmiss bins: $200<\ptmiss\leq350\GeV$, $350< \ptmiss\leq500\GeV$, $\ptmiss>500\GeV$
\end{itemize}
Regions R2 and R4, which have high \MJ, are further divided into bins according to  
the number of \smalr jets (\njets) and the number of \PQb-tagged jets (\nb) as follows:
\begin{itemize}
\item two \njets bins: $6\leq\njets\leq8$, $\njets\geq9$,
\item three \nb bins: $\nb=1$, $\nb=2$, $\nb\geq3$,
\end{itemize}
giving a total of 18 bins each.
Backgrounds with a single \PW~boson decaying leptonically are strongly suppressed by the requirement $\mt>140$\GeV,
so the background in R3 and R4 is dominated by dilepton \ttbar events.
Approximately half of the dilepton background events in R4 contain a missed
electron or muon, and the other half contain a hadronically decaying $\tau$ lepton.
Given that the main background processes have two or fewer \PQb quarks, the total SM contribution to the
$\nb\geq3$ bins is very small and is driven by the \PQb tag misidentification rate.
Signal events in the T1tttt and T5tttt models populate primarily the bins with $\nb\geq 2$.

The method for predicting the background yields takes advantage of the near absence of
correlation between the \MJ and \mt variables in R1--R4, which is a consequence
of the high jet multiplicity, \ptmiss, and \ST requirements applied
in the baseline selection~\cite{Khachatryan:2016uwr}. To satisfy these requirements, background
events must typically contain additional jets from ISR.
Even though the background at low \mt arises largely from single-lepton $\ttbar$ events, while the background at high
\mt is dominated by dilepton $\ttbar$ events,
the shapes of the \MJ distributions at low and high \mt
become very similar in the presence of multiple ISR jets.
We therefore measure this shape
at low \mt (R1, R2) and extrapolate it to high \mt
to obtain the background prediction in R4.
The fitted mean background yields in R1--R4 are thus related by the constraint
${\mu_{\text{R4}}^{\text{bkg}}=\kappa\,\mu_{\text{R3}}^{\text{bkg}}\,\mu_{\text{R2}}^{\text{bkg}}/\mu_{\text{R1}}^{\text{bkg}}}$.
Here, $\kappa$ is a near-unity correction factor obtained from MC simulation of the total background that accounts for a residual
\mt-\MJ correlation:
\begin{equation}
  \kappa = \frac{\mu_{\text{R4}}^{\text{MC bkg}}/\mu_{\text{R2}}^{\text{MC bkg}}}{\mu_{\text{R3}}^{\text{MC bkg}}/\mu_{\text{R1}}^{\text{MC bkg}}}.\label{eq:bkgest:kappa}
\end{equation}
This constraint is imposed by relating the expected yields in R1--R4 to three parameters: an overall
background normalization $\lambda$ and two ratios $R(\mt)$ and $R(\MJ)$, where the expected background yields
are given by $\mu_{\text{R1}}^{\text{bkg}}=\lambda$, $\mu_{\text{R2}}^{\text{bkg}}=\lambda\, R(\MJ),$
$\mu_{\text{R3}}^{\text{bkg}}=\lambda\, R(\mt)$, and
$\mu_{\text{R4}}^{\text{bkg}}=\kappa\,\lambda\,R(\MJ)\,R(\mt)$. 
These quantities are defined such that there is one value of $R(\MJ)$ and $\kappa$ for each bin
of \ptmiss, \njets, and \nb. Because regions R1 and R3 are integrated in \njets and \nb, the fit parameters
$\lambda$ and $R(\mt)$ are defined such that there is only one value of these quantities for each bin
in \ptmiss.

We perform two types of maximum likelihood fits,
which are described in detail in Ref.~\cite{Khachatryan:2016uwr}.
The \textit{predictive fit} uses
the observed yields in R1--R3, assuming no signal contribution,
to propagate the uncertainties to $\lambda$, $R(\MJ)$,
and $R(\mt)$. The \emph{global fit} uses the observed yields in all four regions R1--R4
and allows a signal contribution with a single normalization parameter.
The global fit accounts for signal contamination in R1--R3, which is typically less than 10\%,
and is used to compute signal limits and
significances. The results from the predictive fit simplify theoretical reinterpretation
in terms of other models by only requiring comparison of observed and
predicted yields in R4 rather than all four regions. In both cases, the likelihood
function is written as a product of Poisson distributions for the relevant contributions in
bins of $\ptmiss$, $\njets$, and $\nb$ within R2 and R4, taking into account
the correlated yields between the unbinned regions R1 and R3.

Systematic uncertainties in the background prediction are incorporated in
the uncertainty in the double ratio correction factor $\kappa$.
Discrepancies between the value of $\kappa$ predicted by simulation
and the true value of $\kappa$ in the data can in principle arise from
mismodeling of the background composition or its properties, including
detector effects.

To assess the potential impact of such effects on $\kappa$, two control samples in data
are used: a 5-jet control sample and a dilepton control sample.
The 5-jet control sample is completely dominated by
background processes and has an SM composition very similar to that of the
analysis regions. In particular, this sample probes the
rate at which \ptmiss is mismeasured in single-lepton events, which could increase the tail of the
$\mt$ distribution. Such events account for about 7\% of the background in the signal
region at high \ptmiss. This small event category can have a $\kappa$ value that
departs significantly from unity, and it is important to validate the modeling
of such effects. Using the analogous R1--R4 regions in the $\njets=5$ control sample,
$\kappa$ values are measured in data and are found to be consistent with those obtained from simulation. Because of this
consistency, the statistical uncertainty obtained from the comparison in the
$\njets=5$ control sample is assigned as an uncertainty in $\kappa$ for each
\ptmiss bin. These uncertainties are taken to be fully correlated over the \njets and \nb bins.

The dilepton control sample is used to test
the degree of similarity between the \MJ shapes
of single-lepton and dilepton $\ttbar$
events in the presence of ISR.
This sample includes not only events with two identified isolated leptons,
but also events with one lepton and an oppositely charged veto track.
The usual R3 and R4 regions are replaced by dilepton events,
and the quantity $\kappa$ is measured in bins of
\njets. As in the 5-jets control sample,
the values of $\kappa$ measured in data are found to be consistent
with those observed in simulation, and uncertainties are assigned in a similar way.
The uncertainties are treated as
independent across $\njets$ bins but fully correlated across $\nb$ and $\ptmiss$ bins.
The uncertainties from the dilepton and 5-jet control samples
are treated as uncorrelated. Studies of a broad range of potential mismodeling effects in simulation
show that all such effects would be evident in these control samples.

Systematic uncertainties in the expected signal yields
account for uncertainties in the trigger, lepton identification, jet identification, and b tagging
efficiencies in simulated data; uncertainties in the distributions of \ptmiss,
number of pileup vertices, and ISR jet multiplicity; and uncertainties in the jet energy
corrections, QCD scales, and integrated luminosity~\cite{CMS-PAS-LUM-17-001}.
The combined effect of all signal-related uncertainties is typically about 25\%.

Table~\ref{tab:results:yields} lists the observed event yields in region R4 in
data, together with the mean background yields from the predictive fit and the expected signal
yields from two benchmark model points.
\begin{table}[tbp!]
  \centering
  \topcaption{Observed event yields and mean background yields from the predictive
    fit in the 18 bins of the signal region R4. Each bin is specified by the values
    of $\ptmiss$, \njets, and \nb. The uncertainties in $\kappa$
    include both a statistical component from the size of the MC samples and
    a systematic component assessed from the data control samples. The uncertainty
    in the predicted event yield includes both of these and the statistical
    uncertainties associated with the data control regions. Yields for the two T1tttt benchmark models NC and C are also given.\label{tab:results:yields}}
  \begin{scotch}{llccccc}
    \njets & \nb        & NC         & C   & $\kappa$    & Pred.         & Obs. \\
    \hline
    \multicolumn{7}{c}{$200< \ptmiss\leq 350\GeV$}            \\ \hline
    6-8     & 1        & 0.4       & 1.9 & $1.2\pm0.2$ & $85\pm14$ & 106  \\
    6-8     & 2        & 0.6       & 3.0 & $1.2\pm0.2$ & $55.1\pm9.3$  & 75   \\
    6-8     & $\geq$3  & 0.6       & 2.2 & $1.5\pm0.2$ & $16.4\pm3.0$  & 16   \\
    $\geq$9 & 1        & 0.2       & 1.6 & $1.0\pm0.2$ & $6.5\pm1.5$   & 11   \\
    $\geq$9 & 2        & 0.3       & 2.1 & $1.2\pm0.3$ & $7.6\pm1.9$   & 11   \\
    $\geq$9 & $\geq$3  & 0.4       & 3.1 & $1.4\pm0.3$ & $2.3\pm0.7$   & 2    \\
    \hline
    \multicolumn{7}{c}{$350 < \ptmiss\leq 500\GeV$}            \\ \hline
    6-8     & 1        & 0.7       & 1.1 & $1.0\pm0.3$ & $17.4\pm6.6$  & 25   \\
    6-8     & 2        & 0.9       & 1.3 & $1.1\pm0.4$ & $13.7\pm5.3$  & 10   \\
    6-8     & $\geq$3  & 0.8       & 0.9 & $1.3\pm0.4$ & $3.8\pm1.6$   & 1    \\
    $\geq$9 & 1        & 0.3       & 1.0 & $1.1\pm0.4$ & $1.3\pm0.6$   & 2    \\
    $\geq$9 & 2        & 0.5       & 1.1 & $0.8\pm0.3$ & $1.6\pm0.8$   & 2    \\
    $\geq$9 & $\geq$3  & 0.7       & 2.1 & $1.2\pm0.5$ & $0.6\pm0.4$   & 0    \\
    \hline
    \multicolumn{7}{c}{$\ptmiss>500\GeV$}                          \\ \hline
    6-8     & 1        & 2.5       & 0.6 & $1.0\pm0.3$ & $1.9\pm1.5$   & 8    \\
    6-8     & 2        & 3.6       & 1.0 & $1.0\pm0.4$ & $0.9\pm0.7$   & 4    \\
    6-8     & $\geq$3  & 3.2       & 0.4 & $1.5\pm0.6$ & $0.4\pm0.4$   & 1    \\
    $\geq$9 & 1        & 1.0       & 0.7 & $1.0\pm0.4$ & $0.2\pm0.2$   & 2    \\
    $\geq$9 & 2        & 1.8       & 1.2 & $1.0\pm0.4$ & $0.1\pm0.1$   & 0    \\
    $\geq$9 & $\geq$3  & 2.3       & 1.7 & $3.1\pm1.5$ & $0.1\pm0.1$   & 0    \\
  \end{scotch}
\end{table}
The uncertainties in the predicted background yields include the statistical
uncertainties on the event yields in R1--R4 in data, the statistical uncertainties in the
$\kappa$ values arising from the finite size of simulated event samples, and the
systematic uncertainties in $\kappa$ as assessed from the data control samples.
The observed yields are consistent with
the background predictions in all of the 18 signal bins within 2 standard deviations (s.d.),
with most of the 18 bins consistent within 1 s.d.
The R4 bins with $\ptmiss>500\GeV$ show an underprediction of the background with respect to the observed
yields. However, accounting for the correlations arising from the use of a single, integrated
yield in R3 across bins in $\njets$ and $\nb$, the
significance of the discrepancy in these six bins in R4 is only 1.9 s.d.,
mostly due to the bins with $\nb=1$.

\begin{table}[tbp!]
  \centering
  \topcaption{Observed event yields and mean background yields from the predictive
    fit in four aggregate search bins. In all four cases, the predicted yields
    refer to the signal region R4 with the standard $\mt>140\GeV$ and
    $\MJ>400\GeV$ requirements applied in addition to the baseline selection. Unlike the
    finely binned approach, where all 18 background predictions are computed
    simultaneously, the four aggregate bin predictions are computed
    separately. The aggregate bins overlap, causing their background predictions to be highly correlated.
    Yields for the two T1tttt benchmark models NC and C are also given.
    \label{tab:results:agg}}
    \begin{scotch}{lcccccccc}
      \ptmiss [\GeVns{}] & \njets & \nb  & NC                         & C   & $\kappa$    & Pred.       & Obs. \\
      \hline
      $>$200 & $\geq$9 & $\geq$3 & 3.4                        & 6.9 & $1.4\pm0.3$ & $3.1\pm0.8$ & 2    \\
      $>$350 & $\geq$9 & $\geq$2 & 5.3                        & 6.2 & $1.0\pm0.4$ & $2.7\pm1.2$ & 2    \\
      $>$500 & $\geq$6 & $\geq$3 & 5.4                        & 2.1 & $1.7\pm0.6$ & $0.5\pm0.4$ & 1    \\
      $>$500 & $\geq$9 & $\geq$1 & 5.1                        & 3.6 & $1.2\pm0.4$ & $0.4\pm0.4$ & 2    \\
    \end{scotch}
\end{table}

To simplify the reinterpretation of the results in terms of other theoretical models, we provide predicted
mean background yields for four aggregated search bins, shown in Table~\ref{tab:results:agg}.  The aggregate
bins are defined such that at least one bin will provide sensitivity to most of the models for which the
finely binned analysis has sensitivity. Since the aggregate bins overlap, they are intended to be used one at
a time, unlike the 18 nonoverlapping signal bins, which are considered simultaneously in the fit. Each
prediction includes all sources of uncertainty. The choice of the best aggregate bin will depend on the model
under study. For the T1tttt benchmark models considered in this letter, using the aggregate bins results
in expected upper limits on the cross sections that are 20--50\% higher than those 
resulting from the full analysis.

\begin{figure}[tbh]
  \centering
  \includegraphics[width=\cmsFigWidthii]{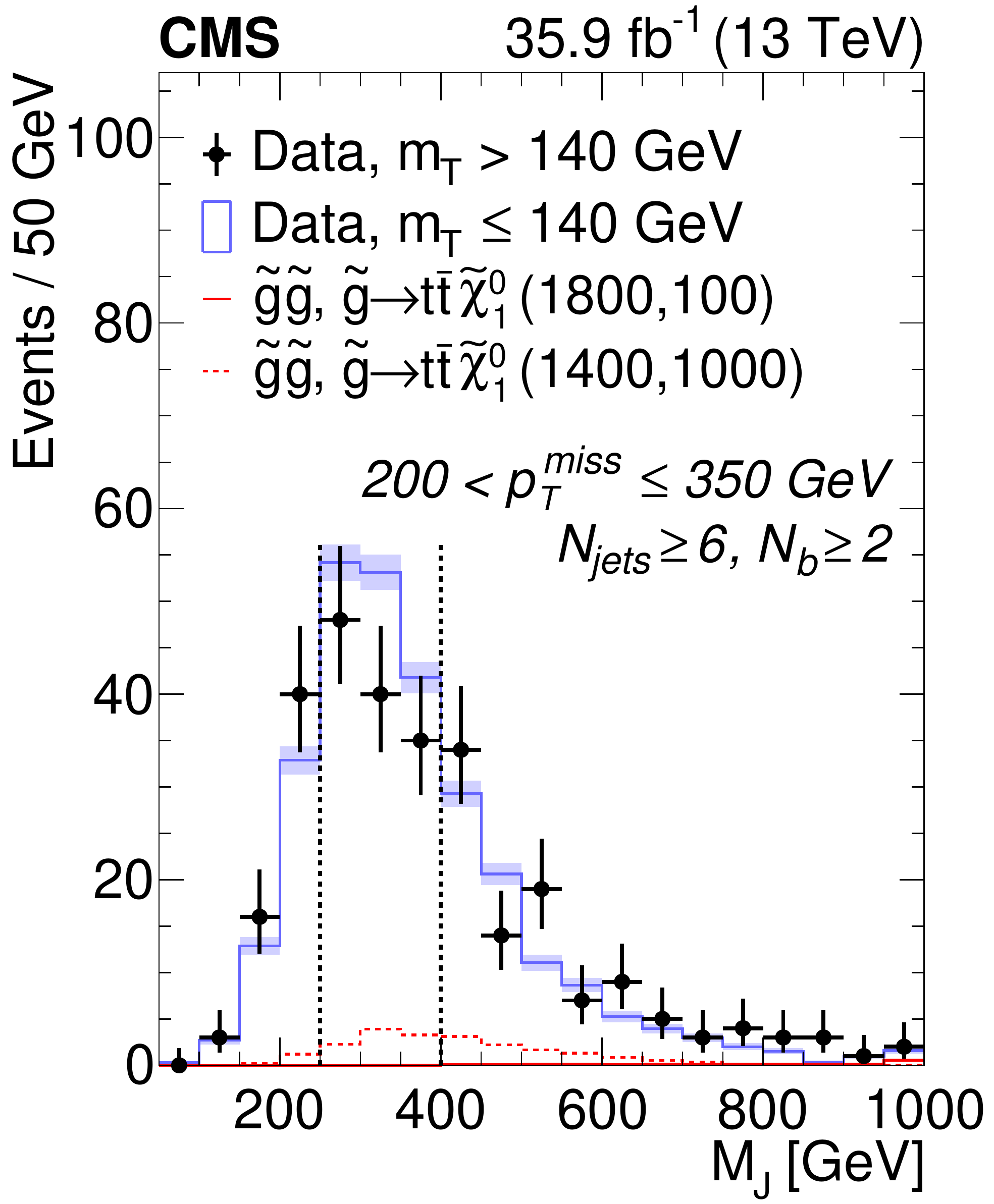}
  \includegraphics[width=\cmsFigWidthii]{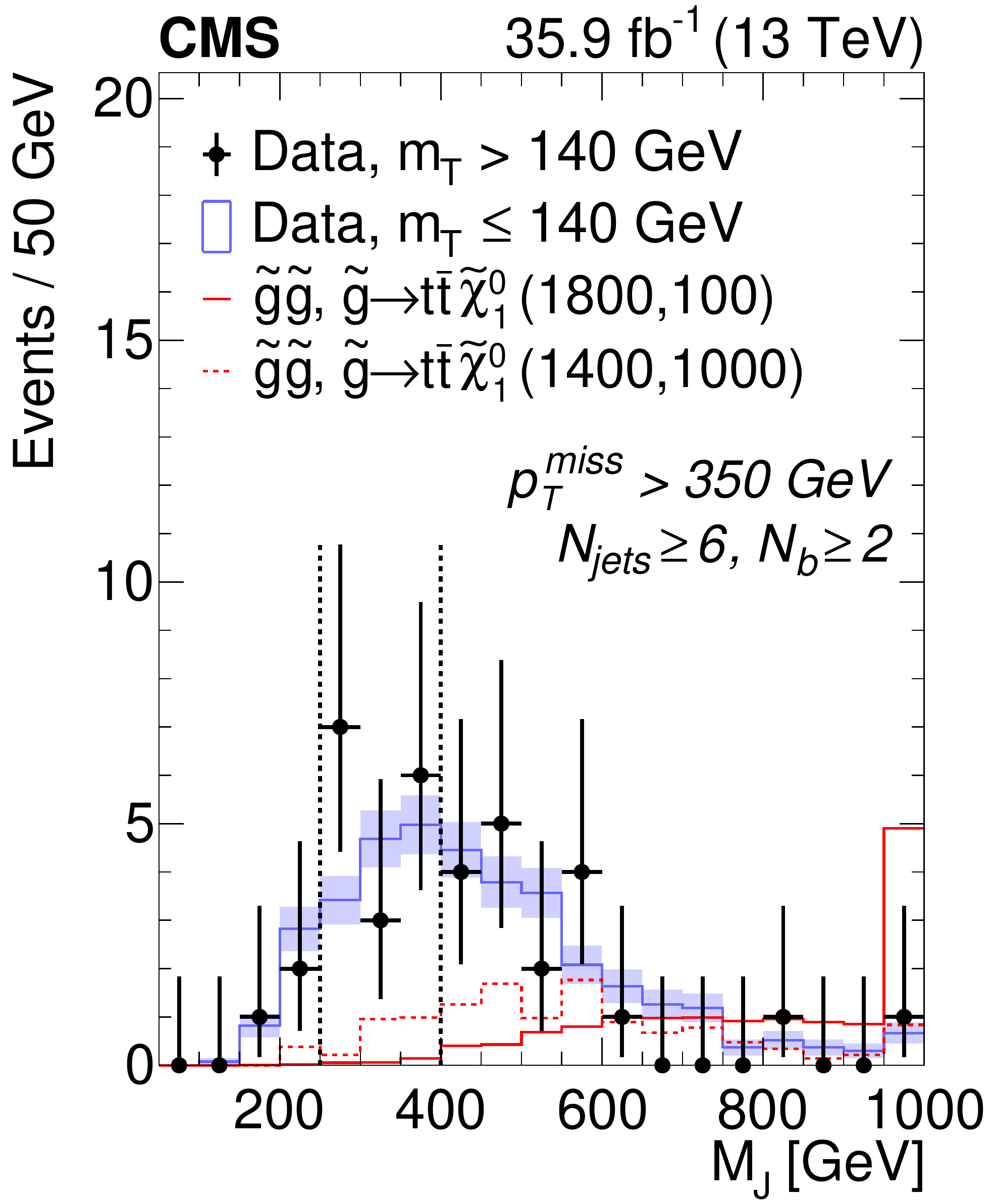}
  \caption{Distributions of \MJ observed in data for $200<\ptmiss\leq 350$\GeV
    (left) and $\ptmiss>350$\GeV (right) with the baseline selection and either $\mt\leq
    140$\GeV or $\mt>140$\GeV. In each plot, the data at
    low $\mt$ have been normalized to the yield observed at high \mt.
    The vertical dashed lines at $\MJ=250$\GeV and 400\GeV show the boundaries separating the
    control and signal regions. The data are integrated over $\njets\geq 6$ and $\nb \geq 2$. Two SUSY
    benchmark models, whose contributions are small in the lower $\ptmiss$ region,
    are shown in the solid and dashed red histograms. Overflow
    events are included in the uppermost bins.
}
  \label{fig:results:MJ}
\end{figure}

Figure~\ref{fig:results:MJ} compares the shapes of the $\MJ$ distributions observed in data in the
single-lepton sample for $\mt\leq 140$\GeV and $\mt>140$\GeV in two regions of \ptmiss.  The shapes of the
two \MJ distributions for each \ptmiss region are very similar, as expected in the absence of signal. A
further correction is applied via the $\kappa$ factors listed in Table~\ref{tab:results:yields} in \MJ ranges
larger than the binning shown in the figure.  The lower-\ptmiss region shows the background behavior with
higher statistics, while the higher-\ptmiss region has higher sensitivity to the signal.

\begin{figure}[tbh]
  \centering
  \includegraphics[width=\cmsFigWidth]{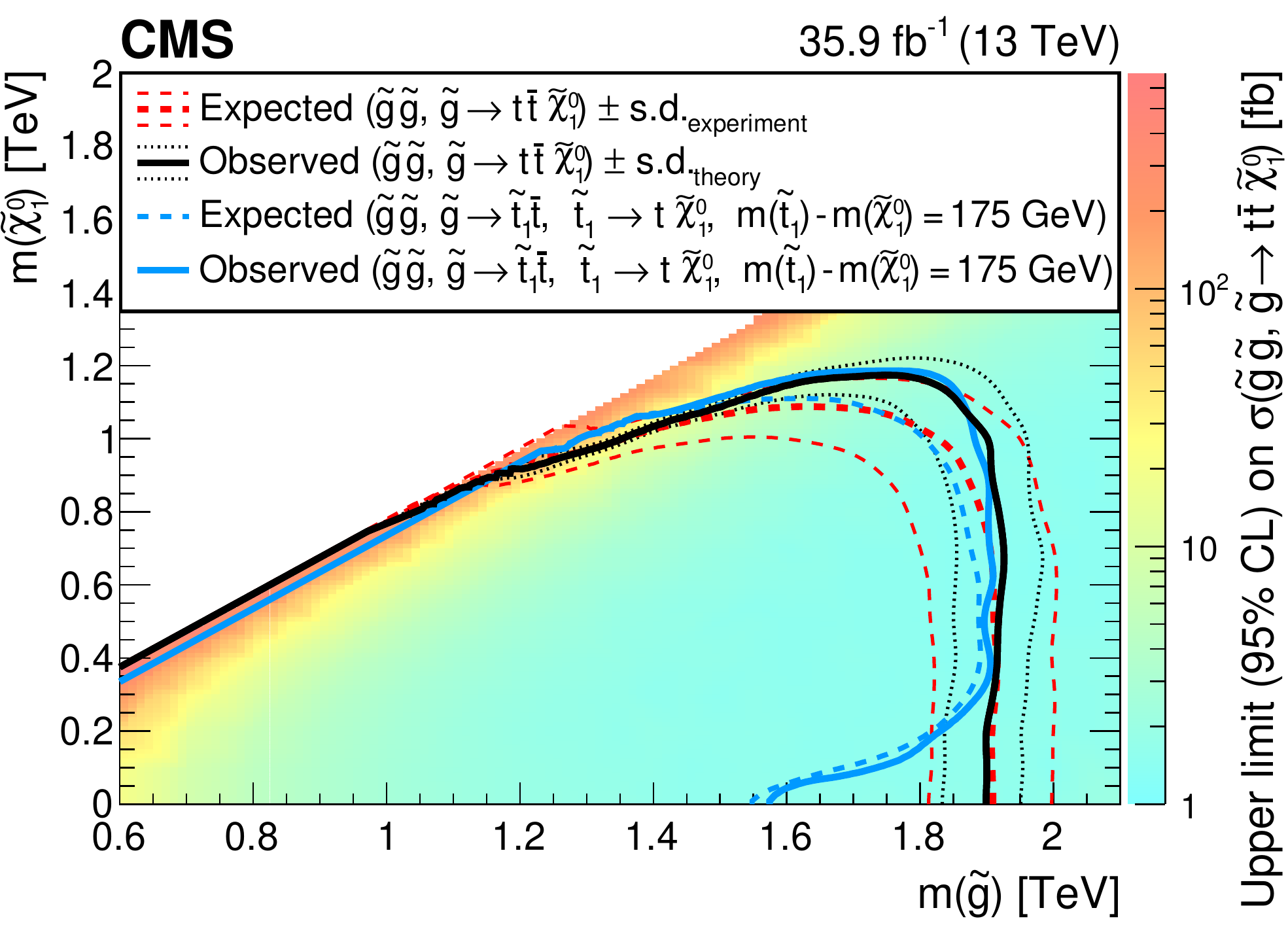}
  \caption{Observed and expected excluded gluino and neutralino masses at 95\%~\CL for the T1tttt and T5tttt
    models. The black (red) lines show the observed (expected) exclusion and the range associated
    with the theoretical (experimental) uncertainties for the T1tttt model.  The solid (dashed) blue line
    shows the observed (expected) exclusion for the T5tttt model. The uncertainties for the T5tttt exclusion
    limits are not shown and are similar to those for the T1tttt model.  The color map shows the observed
    cross section upper limits for the T1tttt model.
    } \label{fig:results:interp}
\end{figure}

Figure~\ref{fig:results:interp} shows an interpretation of the results as exclusion limits at 95\% confidence
level (\CL) for T1tttt and T5tttt.  The limits are obtained using the CL$_\mathrm{s}$ method with a
profile-likelihood ratio as the test statistic, using asymptotic approximations for the distribution of the
test statistic~\cite{0954-3899-28-10-313,CMS-NOTE-2011-005,Cowan:2010js}.  The color map shows the cross
section upper limits as a function of $\mGlu$ and $\mLSP$ for T1tttt, assuming a 100\% branching fraction for
the decay $\PSg\to\ttbar\PSGczDo$. The T1tttt model points below the dark solid curve, which extend up to
gluino masses of about 1.9\TeV for neutralino masses up to 1\TeV, have a theoretical cross section above the
observed cross section upper limit and are thus excluded by this analysis.  The dotted black lines around the
observed mass limits show the impact of the theoretical uncertainties in the overall signal cross sections
arising from uncertainties in the parton distribution functions and the renormalization and factorization
scales.

Model points below the light solid curve are excluded at 95\%~\CL for the T5tttt model, where it is assumed
that the top squark mass is 175\GeV above the neutralino mass, a limiting case in terms of sensitivity to the
decay kinematics. The T5tttt simulation does not explicitly include direct top squark pair production. Studies
presented in Ref.~\cite{Khachatryan:2016uwr} demonstrate that the effect of this contribution is very small
for most of the space of T5tttt model points considered here.  For most of the excluded region, the boundaries
for T1tttt and T5tttt are very similar, indicating only a weak overall sensitivity to the value of the top
squark mass. At low values of $\mLSP$ in T5tttt, the sensitivity is reduced because the neutralino carries
very little momentum; however, some sensitivity is still provided by dilepton events that escape the lepton
veto~\cite{Khachatryan:2016uwr}.  For both the T1tttt and T5tttt models, expected limits are computed using
the background-only hypothesis, with nuisance parameters assuming their best fit values from the observed
data. All limits are computed using results from the global fit.

In summary, we have performed a search for an excess event yield above that expected for standard model (SM)
processes using a data sample of proton-proton collision events with an integrated luminosity of 35.9\fbinv at
$\sqrt{s} = 13$\TeV. The signature is characterized by large missing transverse momentum, a single isolated
lepton, multiple jets, and at least one $\PQb$-tagged jet.  No significant excesses above the SM backgrounds
are observed.  The results are interpreted in the framework of simplified models that describe natural
supersymmetry (SUSY) scenarios.  For gluino pair production followed by the three-body decay
$\PSg\to\ttbar\PSGczDo$ (T1tttt model), gluinos with masses below 1.9\TeV are excluded at 95\% confidence
level for neutralino masses up to about 1\TeV.  For the two-body gluino decay $\PSg\to\PSQt_1\cPaqt$ with
$\PSQt_1\to\cPqt\PSGczDo$ (T5tttt model), the results are generally similar, except at low neutralino masses,
where the excluded gluino mass is somewhat lower.  These results extend previous gluino mass limits by about
300\GeV and are among the most stringent constraints on these simplified models of SUSY to date.

\begin{acknowledgments}
We congratulate our colleagues in the CERN accelerator departments for the excellent performance of the LHC and thank the technical and administrative staffs at CERN and at other CMS institutes for their contributions to the success of the CMS effort. In addition, we gratefully acknowledge the computing centers and personnel of the Worldwide LHC Computing Grid for delivering so effectively the computing infrastructure essential to our analyses. Finally, we acknowledge the enduring support for the construction and operation of the LHC and the CMS detector provided by the following funding agencies: BMWFW and FWF (Austria); FNRS and FWO (Belgium); CNPq, CAPES, FAPERJ, and FAPESP (Brazil); MES (Bulgaria); CERN; CAS, MoST, and NSFC (China); COLCIENCIAS (Colombia); MSES and CSF (Croatia); RPF (Cyprus); SENESCYT (Ecuador); MoER, ERC IUT, and ERDF (Estonia); Academy of Finland, MEC, and HIP (Finland); CEA and CNRS/IN2P3 (France); BMBF, DFG, and HGF (Germany); GSRT (Greece); OTKA and NIH (Hungary); DAE and DST (India); IPM (Iran); SFI (Ireland); INFN (Italy); MSIP and NRF (Republic of Korea); LAS (Lithuania); MOE and UM (Malaysia); BUAP, CINVESTAV, CONACYT, LNS, SEP, and UASLP-FAI (Mexico); MBIE (New Zealand); PAEC (Pakistan); MSHE and NSC (Poland); FCT (Portugal); JINR (Dubna); MON, RosAtom, RAS, RFBR and RAEP (Russia); MESTD (Serbia); SEIDI and CPAN (Spain); Swiss Funding Agencies (Switzerland); MST (Taipei); ThEPCenter, IPST, STAR, and NSTDA (Thailand); TUBITAK and TAEK (Turkey); NASU and SFFR (Ukraine); STFC (United Kingdom); DOE and NSF (USA).
\end{acknowledgments}
\bibliography{auto_generated}
\cleardoublepage \appendix\section{The CMS Collaboration \label{app:collab}}\begin{sloppypar}\hyphenpenalty=5000\widowpenalty=500\clubpenalty=5000\input{SUS-16-037-authorlist.tex}\end{sloppypar}
\end{document}

%% file: SUS-16-037-authorlist.tex
\textbf{Yerevan Physics Institute,  Yerevan,  Armenia}\\*[0pt]
A.M.~Sirunyan, A.~Tumasyan
\vskip\cmsinstskip
\textbf{Institut f\"{u}r Hochenergiephysik,  Wien,  Austria}\\*[0pt]
W.~Adam, F.~Ambrogi, E.~Asilar, T.~Bergauer, J.~Brandstetter, E.~Brondolin, M.~Dragicevic, J.~Er\"{o}, M.~Flechl, M.~Friedl, R.~Fr\"{u}hwirth\cmsAuthorMark{1}, V.M.~Ghete, J.~Grossmann, J.~Hrubec, M.~Jeitler\cmsAuthorMark{1}, A.~K\"{o}nig, N.~Krammer, I.~Kr\"{a}tschmer, D.~Liko, T.~Madlener, I.~Mikulec, E.~Pree, D.~Rabady, N.~Rad, H.~Rohringer, J.~Schieck\cmsAuthorMark{1}, R.~Sch\"{o}fbeck, M.~Spanring, D.~Spitzbart, J.~Strauss, W.~Waltenberger, J.~Wittmann, C.-E.~Wulz\cmsAuthorMark{1}, M.~Zarucki
% ...
\vskip\cmsinstskip
% ...
1:~~Also at Vienna University of Technology, Vienna, Austria\\
% ...
